\documentclass[aps,prd,10pt,twocolumn,nofootinbib,groupedaddress,amsfonts,floatfix]{revtex4-1}

\usepackage[colorlinks=true,urlcolor=blue,linkcolor=,citecolor=blue]{hyperref}
\usepackage{amsmath,amssymb,amstext,amsbsy,amsfonts,amsthm,graphicx,microtype,dsfont}
\usepackage[capitalize]{cleveref}

\creflabelformat{equation}{(#2\textup{#1}#3)}

\newcommand{\prg}{\mathrm{pr}}
\newcommand{\mpl}{M_\mathrm{Pl}}
\newcommand{\op}{\Omega_\mathrm{p}}
\newcommand{\rv}{r_\mathrm{v}}

\newcommand{\diff}[1]{\mathrm{d}{#1}\,}

\def\be{\begin{equation}}
\def\ee{\end{equation}}
\def\ba{\begin{eqnarray}}
\def\ea{\end{eqnarray}}

\newcommand{\nn}{\nonumber}

\renewcommand{\d}{\mathrm{d}}

\newcommand{\eg}{e.g.}

\newcommand{\eqn}[1]{eq.~#1}

\usepackage{color}
\usepackage{ifthen}
\newboolean{editorial}
\setboolean{editorial}{true}
\newcommand{\editorial}[2]{\ifthenelse{\boolean{editorial}}{\textcolor{red}{[\textsf{\textbf{{#1}}}: }\textcolor{blue}{\textsf{{#2}}}\textcolor{red}{]}}{}}

\usepackage{xcolor}

\begin{document}

\title{A Well-Posed UV Completion for Simulating Scalar Galileons}

\author{Mary Gerhardinger${}^1$}
\author{John T. Giblin, Jr${}^{1,2}$}
\author{Andrew J. Tolley${}^{3,2}$}
\author{Mark Trodden${}^4$}
\affiliation{${}^1$Department of Physics, Kenyon College, Gambier, Ohio 43022, USA}
\affiliation{${}^2$CERCA/ISO, Department of Physics, Case Western Reserve University, Cleveland, Ohio 44106, USA}
\affiliation{${}^3$Theoretical Physics, Blackett Laboratory, Imperial College, London, SW7 2AZ, UK}
\affiliation{${}^4$Center for Particle Cosmology, Department of Physics and Astronomy, University of Pennsylvania, Philadelphia, Pennsylvania 19104, USA}

\email{gerhardinger1@kenyon.edu \\ giblinj@kenyon.edu \\ a.tolley@imperial.ac.uk \\ trodden@physics.upenn.edu}

\begin{abstract}
The Galileon scalar field theory is a prototypical example of an effective field theory that exhibits the Vainshtein screening mechanism, which is incorporated into many extensions to Einstein gravity.  The Galileon describes the helicity zero mode of gravitational radiation, the presence of which has significant implications for predictions of gravitational waves from orbiting objects, and for tests of gravity sensitive to additional polarizations. Because of the derivative nature of their interactions, Galileons are superficially not well-posed as effective field theories. Although this property is properly understood merely as an artifact of the effective field theory truncation, and is not theoretically worrisome, at the practical level it nevertheless renders numerical simulation highly problematic. Notwithstanding, previous numerical approaches have successfully evolved the system for reasonable initial data by slowly turning on the interactions. We present here two alternative approaches to improving numerical stability in Galileon numerical simulations. One of these is a minor modification of previous approaches, which introduces a low pass filter that amounts to imposing a UV cutoff together with a relaxation method of turning on interactions. The second approach amounts to constructing a (numerical) UV completion for which the dynamics of the high momentum modes is under control, and for which it is unnecessary to slowly turn on nonlinear interactions.  We show that numerical simulations of the UV theory successfully reproduce the correct Galileon dynamics at low energies, consistent with the low-pass filter method and with previous numerical simulations.
\end{abstract}

\maketitle


\section{Introduction}

In the current era of high precision cosmology and gravitational wave physics, there is a significant interest in understanding how this precise data can be used to test our theories of gravity and low energy particle physics, and to search for potential new physics beyond the standard models. In order to do this we need to have a detailed understanding of the predictions of extensions to standard theories. Effective field theories are a natural tool to describe these corrections, and in the last decade many interesting effective field theories have been developed that extend Einstein gravity while successfully reproducing its successful predictions. 
A large class of theoretical models that have been developed as alternative descriptions of dark energy and late time acceleration are those that incorporate the Vainshtein screening mechanism \cite{Vainshtein:1972sx} (see \cite{Babichev:2013usa} for a review).
This mechanism is built into massive theories of gravity \cite{Deffayet:2001uk,Dvali:2002vf,Lue:2003ky,Babichev:2009us,Babichev:2009jt,Babichev:2010jd,deRham:2010ik,deRham:2010kj} and allows them to be made consistent with solar system tests of gravity by screening the would-be fifth forces that are propagated by the additional helicity-zero mode of the massive graviton. While the complete theory of massive gravity is quite complex \cite{deRham:2010kj}, these essential features are captured by a simplified scalar field theory, the Galileon \cite{Nicolis:2008in}, which incorporates the nonlinear interactions that are responsible for the screening mechanism \cite{deRham:2010ik,Ondo:2013wka}.  

The nonlinear nature of the Vainshtein screening mechanism means that it is difficult to describe analytically beyond very special configurations, and the traditional post-Newtonian or post-Minkowskian formalisms fail to adequately describe the essential features.  Given this, numerical progress is crucial, but is unfortunately hampered by the fact that the Galileon effective theory is not well-posed, and despite admitting second order equations of motion it does have regimes in which the equations are no longer hyperbolic. However, such regions arise only when the interactions are taken to be large, for which it is unclear that the effective field theory is under control. Thus, while not a fundamental theoretical problem, successful numerical approaches need to incorporate a mechanism through which to avoid these dangerous regions. For example, in \cite{Dar:2018dra} the cubic Galileon was successfully simulated by slowly turning on the non-linear interactions to avoid instabilities. 

In this paper we propose a different scheme which replaces the original Galileon theory with a well-posed (numerical) UV completion by means of the introduction of auxiliary higher spin fields. This approach is similar in spirit to that proposed in \cite{Cayuso:2017iqc,Allwright:2018rut}, based on the  M\"uller-Israel-Stewart formulation \cite{muller1967paradoxon,ISRAEL1976213,ISRAEL1979341,ISRAEL1976310}\footnote{An alternative numerical scheme is to construct the solution perturbatively in the EFT corrections as in \cite{Okounkova:2017yby,Okounkova:2019dfo}. Issues with secular growth of such a perturbative expansion can potentially be resummed in the manner proposed in \cite{GalvezGhersi:2021sxs}. We do not consider these approaches as the Vainshtein screening region is necessarily non-perturbative in the leading EFT derivative interactions.}. The key difference is that here the additional spin 1 and spin 2 fields will be given propagating (hyperbolic) equations in a manner which is closely motivated by the massive spin 2 origin of the Galileon. The cubic Galileon arises consistently as the leading terms in the low energy effective theory of our proposed UV completion, and so we anticipate that a successful numerical treatment will correctly reproduce the dynamics of the Galileon at long wavelengths. We stress that the UV completion here is a numerical one, since no Lorentz invariant local and unitary UV completion of the Galileon is known, and there are now strong arguments suggesting that one does not exist \cite{Tolley:2020gtv}.\footnote{These arguments hinge crucially on locality and Lorentz invariance. If the UV theory is mildly non-local then there may be no problem \cite{Keltner:2015xda}.  In addition fractons are a Lorentz violating realization of Galileons \cite{Pretko:2018jbi,Seiberg:2019vrp}} Since our goal here is to render the low energy theory numerically well defined, we are not constrained by the need to find a local Lorentz invariant action, and thus we propose only a local UV extension of the equations of motion which reproduces the Galileon at low energies. Our approach should render the theory well-behaved when simulating any type of physics, but we will focus mostly on solutions relevant to radiation generated by orbiting binary objects, since this is where the most immediately interesting numerical applications are likely to be.


\section{Cubic Galileon}

We begin by looking at the action for the cubic Galileon (see \eg \cite{deRham:2012fw,deRham:2012fg})
\begin{align}
\label{eqn:galaction}
S = \int \diff{^4 x}\left(-\frac34 (\partial\pi)^2-\frac1{4\Lambda^3} (\partial\pi)^2 \Box \pi + \frac{1}{2\mpl}\pi T \right)\,,
\end{align}
where $T$ is the trace of the stress-energy tensor for the matter content. Note that we use a non-standard choice of normalization and coupling that is consistent with how the Galileon degree of freedom arises as the helicity-zero mode in theories of gravity where the graviton is effectively massive.
This action yields a classical equation of motion,
\begin{equation}
\label{eqn:eom}
\Box\pi +\frac{1}{3\Lambda^3}\left(  (\Box\pi)^{2} - (\partial_\mu\partial_\nu\pi)^2 \right) = -\frac{T}{3 \mpl},
\end{equation}
which makes manifest the Galileon symmetry $\pi \rightarrow \pi + c+ v_{\mu} x^{\mu}$, and in which the nonlinearity is parameterized by the strength of the coupling, $1/3\Lambda^3$.

In the case where the source is spherically-symmetric and time-independent, with an associated source mass $M$, there exist static, analytic solutions to \eqn\eqref{eqn:eom},
\begin{equation}
E(r) = -\frac{\Lambda^3}{4}r\left[3-\sqrt{9+\frac{32}{\pi}\left(\frac{\rv}{r}\right)^3}\right],
\end{equation}
where $E\equiv \partial \pi/\partial r$, and where we have defined the Vainshtein radius,
\begin{equation}
\rv \equiv \frac1{\Lambda}\left(\frac{M_{\rm s}}{16 \mpl}\right)^{1/3} \, .
\end{equation}
The Vainshtein radius sets the distance from the center of mass of the source at which the nonlinear interactions of the Galileon become important. A key feature of the Vainshtein screening mechanism is that this distance is astrophysically large, meaning that most dynamical systems, such as binary pulsars, lie well inside their own Vainshtein radius. For this reason, all linear or perturbative approaches fail to describe the physics of these systems.

This system was studied numerically in \cite{Dar:2018dra}, where it was shown that \eqn\eqref{eqn:galaction} could be simulated using numerical tools \cite{Child:2013ria}.  This work showed that the scalar power radiated by this system followed anticipated scaling relationships.  While this was an important proof of concept, the {\sl numerical} challenges of simulating this system for more realistic hierarchies fundamentally arise from the physical system itself.  When setting up numerical simulations, it is the normal practice to chose numerical parameters to place the physics of greatest interest `well inside' the box. However, in situations where there are many physical scales, this becomes a more difficult problem. Nevertheless, if the system is formally well-posed, then any UV dynamics of the system remains decoupled from the IR physics of interest. When the system is not manifestly well-posed (as happens, for example, when including higher-derivative operators that inevitably arise from quantum corrections within all EFTs (for a detailed discussion see~\cite{Solomon:2017nlh})), then this decoupling of the UV and IR modes is no longer guaranteed. Although effective field theorists have well-established analytic techniques for handling this behavior, it can cause serious problems for numerical implementations.
In particular, in our system, if these modes become populated, they run the risk of violating the assumptions of the effective field theory and are {\sl analytically} unstable. 

The question, then, is whether we can regulate these higher-frequency modes using either numerical techniques to dampen them or by finding a physical UV-completion.  In this work we compare these two techniques by developing examples of both, and studying whether they are stable, and provide solutions that are consistent with a full-numerical solution to \eqn\eqref{eqn:eom}.

\subsection{The UV-completion}

In standard effective field theories, higher dimension operators can be understood as arising from integrating out high energy (UV) physics, either via tree level or loop level effects. For example, when there exists a `tree level' UV completion, this means that it is possible to find an action for a well defined classical UV theory for which explicitly solving the equations of motion for the heavy fields in terms of the light fields as a derivative expansion, and substituting back in the action will result in the action for the desired low energy effective field theory. Since the would-be UV completion is valid at arbitrary high energy scales, we would expect it to be well posed. Indeed if the UV completion is Lorentz invariant, we would expect the characteristics of the UV theory to match the Lorentz lightcone, which is to say that the front velocity of propagating modes should be luminal.

In practice we are rarely lucky enough to know the UV theory and in many cases it may be possible to argue that one does not exist, at least satisfying familiar principles. In the particular case of the massless Galileon \cite{Adams:2006sv} or massive/weakly-broken Galileon \cite{deRham:2017imi,Tolley:2020gtv}, there are now well established arguments from positivity bounds that appear to rule out a standard local Lorentz invariant UV completion. It should be stressed, however, that there are implicit assumptions in these arguments which are not required of a UV completion (the UV completion may for example be mildly non-local \cite{Keltner:2015xda}), and so this does not rule out the Galileon playing a role as an interesting effective field theory. In particular Lorentz violating Galileons emerge in the context of fractons \cite{Pretko:2018jbi,Seiberg:2019vrp}. They also seem to play a special role in scattering amplitude methods both for Lorentz invariant theories \cite{Cheung:2014dqa,Cachazo:2014xea} and non-relativistic theories \cite{Mojahed:2022nrn,Mojahed:2021sxy}. 

In the present context, our goal is not to find a UV completion satisfying all the principles of unitarity and analyticity, but rather the more modest goal of a completion with high energy behavior that is numerically more stable than that of the initial system \eqref{eqn:eom}. Given this, we do not require an action, and at the price of a mild breaking of Lorentz invariance can introduce friction terms to tame unphysical modes.  Our proposed method is motivated by how the Galileon arises in massive gravity theories as the helicity zero mode of a massive spin 2 field. In particular given a spin-2 field $H_{\mu\nu}$, the helicity-zero part of it is encoded in $H_{\mu\nu} \sim \partial_{\mu} \partial_{\nu} \pi$. Indeed, in massive theories of gravity, this enters explicitly via a dynamical gauge transformation $x^{\mu} \rightarrow x^{\mu} + A^{\mu}$ with $A_{\mu} \sim \partial_{\mu} \pi$~\cite{Arkani-Hamed:2002bjr,deRham:2010ik,deRham:2010kj,deRham:2014zqa}. With this in mind, we introduce an auxiliary massive spin-1 field $A_{\mu}$ and an auxiliary massive spin-2 field $H_{\mu\nu}$ that satisfy damped hyperbolic sourced equations. The problematic derivative terms in the Galileon equation of motion are replaced by interactions build out of algebraic functions of the massive spin-2 field. Thus, the UV theory is defined by
\begin{align}
&\Box \pi + \frac{1}{3\Lambda^3}\left(H^{\mu \nu}H_{\mu \nu} - \left(H^\nu_\nu\right)^2\right) = -\frac{T}{3 \mpl} \label{Piequation}\\\
&\Box A_\mu - \frac{1}{\tau} \partial_t A_\mu - M^2 A_\mu = - M^2 \partial_\mu \pi \label{Aequation}\\
&\Box H_{\mu\nu} - \frac{1}{\tau} \partial_t H_{\mu\nu} - M^2 H_{\mu\nu} = - \frac{M^2}{2}\left(\partial_\mu A_\mu + \partial_\nu A_\mu\right).\label{Hequation}\
\end{align}
The presence of the friction terms, parametrized by $\tau^{-1}$, ensures that the homogenous spin-1 and spin-2 mode solutions of \eqref{Aequation} and \eqref{Hequation} decay in a time of order $\tau$. The sources on the RHS of  \eqref{Aequation} and \eqref{Hequation} are introduced to ensure that the particular solutions asmptote at low energies $k,\omega \ll M$ to 
\be
A_{\mu} \sim \partial_{\mu} \pi \label{Aapprox}
\ee
and
\be
H_{\mu\nu} \sim \frac{1}{2}\left(\partial_\mu A_\mu + \partial_\nu A_\mu\right) \sim \partial_{\mu} \partial_{\nu} \pi \, . \label{Happrox}
\ee
Assuming the approximate validity of \eqref{Aapprox} and \eqref{Happrox}, then it is simple to see that \eqref{Piequation} reduces to \eqref{eqn:eom}, which ensures a faithful UV extension. It is apparent that the UV completion \eqref{Piequation},\eqref{Aequation},\eqref{Hequation} has conventional second order equations of motion with characteristics at high energy determined by the Minkowski lightcone. While not a guarantee of stability of the system, this removes the particular problems associated with the derivative interactions present in the Galileon equations of motion \eqref{eqn:eom}. This comes at the cost of replacing the original single field system with a system of 15 dynamical fields. Crucially though, the additional degrees of freedom, even if initially excited, decay away over a time scale $\tau$.

There is an alternative way to write the UV completion that makes its connection with the IR theory more transparent. Assuming the homogenous modes of $H_{\mu \nu}$ and $A_{\mu}$ are set to zero initially, then we may solve for them directly via
\ba
&& H_{\mu \nu} (x) = M^2 \int \d^4 y \, D_{\rm ret}(x,y)   \frac{1}{2}\left(\partial_\mu A_\mu(y) + \partial_\nu A_\mu(y)\right) \, , \nn \\
&& A_{\mu}(x) = M^2 \int \d^4 y \, D_{\rm ret}(x,y)   \partial_\mu \pi(y) \, ,
\ea
where $D_{\rm ret}(x,y)$ is the retarded Green's function satisfying
\be
[\Box - \frac{1}{\tau} \partial_t  - M^2 ] D_{\rm ret}(x,y)=\delta^4(x,y) \, ,
\ee
with solution
\be
D_{\rm ret}(x,y) =\theta(x^0 - y^0) e^{-\frac{(x^0-y^0)}{\tau}} G_{\rm ret}(x,y;M^2-\frac{1}{4 \tau^2}) \, ,
\ee
where $G_{\rm ret}(x,y;M^2-\frac{1}{4 \tau^2})$ is the conventional retarded Green's function for a Klein-Gordon field of mass squared $M^2-\frac{1}{4 \tau^2}$.
Combining these relations we have
\be
H_{\mu \nu} (x) = \int \d^4 y [D^2]_{\rm ret}(x,y) \partial_{\mu} \partial_{\nu} \pi(y) \, , \label{Hnonlocal}
\ee
where
\be
 [D^2]_{\rm ret}(x,y) = \int \frac{\d^4 k}{(2 \pi)^4} \frac{e^{ik.(x-y)} M^4}{(k^2+\frac{i}{\tau} k_0+M^2)^2} \, ,
\ee
vanishes for $x^0-y^0<0$, given that the poles lie in the upper half $k_0$ plane. Substituting \eqref{Hnonlocal} in \eqref{Piequation} yields a causal integro-differential equation for a single degree of freedom $\pi$. At low energies $|k| \ll M^2$ it is apparent that
\be
 [D^2]_{\rm ret}(x,y) \approx \int \frac{\d^4 k}{(2 \pi)^4} e^{ik.(x-y)}  \approx \delta^4(x-y)\, .
\ee
which shows that the leading term in the EFT expansion reproduces the original system \eqref{eqn:eom}.


\section{Numerical Simulations}

From a practical standpoint, the full system described by \eqn\eqref{eqn:eom} is difficult to study numerically due to the facts that: (1) the system is highly non-linear and derivatively coupled, (2) the system has a number of relevant scales that need to be simultaneously resolved and (3) the effective metric for perturbations can become singular \cite{Brito:2014ifa}. Nonetheless, it has been shown that the full system can be simulated numerically, with results that are consistent with analytic estimates.  

To work with the system numerically, we define dimensionless variables, $x^\mu = x^\mu_\prg \bar{r}/2$, $\pi_\prg = \pi \sqrt{\bar{r}/M_{\rm s}}$ to rewrite \eqref{eqn:eom} as
\begin{equation}
\Box_\prg \pi_\prg + \kappa \left(  \left(\Box_\prg \pi_\prg\right)^{2} - \left(\partial^\prg_\mu\partial^\prg_\nu\pi_\prg\right)^2 \right) = -J_\prg
\end{equation}
where
\begin{equation}
\kappa = \frac{1}{3\Lambda} \sqrt{\frac{16M_{\rm s}}{\bar{r}^5}} = \frac{1}{3} r_v^3\left(\frac{16m_{\rm pl}}{M_{\rm s}}\right)\sqrt{\frac{16M_{\rm s}}{\bar{r}^5}}
\end{equation}
is a dimensionless parameter that sets the size of the nonlinear terms and $J_\prg$ is the source.

Throughout our work here, we will focus on the system described in \cite{Dar:2018dra} (which is a numerical implementation of the binary system studied in \cite{deRham:2012fw,Chu:2012kz}) and, as we have commented, will also focus on the dynamics of an orbiting binary system.  To parameterize this system, we generally use two dimensionless quantities: $\beta \equiv \bar{r}/r_v$, which relates the diameter of the orbit (roughly the size of the source) to the Vainshtein radius, and $\alpha \equiv \Omega \bar{r}$, which parameterizes the rotational speed of the system. Roughly speaking, $\beta$ sets the overall mass of the binary system and $\alpha$ sets the strength of the non-linear effects.  With these,
Kepler's Law tells us that 	
	\begin{equation}
		\Omega^2 = \frac{M_{\rm s}}{8 \pi M_{pl}^2 \bar{r}^3} \, ,
	\end{equation}
which fully constrains the system.  In practical terms, this sets  
	\begin{equation}
		\kappa = \frac{32}{2 \sqrt{2 \pi}} \beta^{-3} \alpha ^{-1}.
	\end{equation}
For our fiducial model here, we take $\alpha = 0.2, \beta = 0.05$, which leads to a value of $\kappa \approx 1.70\times 10^5$.

We expect \cite{deRham:2012fw,deRham:2012fg,Chu:2012kz,Dar:2018dra} the system (a Galileon with a cubic interaction) to emit radiation in the quadrupole, with power given by
\begin{equation}
	P_2^\text{cubic}=\frac{M_{\rm s}^2}{8 \pi \mpl^2} \frac{45\times 3^{1/4} \pi^{3/2}}{1024\, \Gamma
   \left(\frac{9}{4}\right)^2} \frac{(\op\bar r)^3}{ (\op\rv)^{3/2}}\op^{2} \, ,
   \label{powerradiated}
\end{equation}
which we can express in a dimensionless way as
\begin{equation}
\frac{\bar{r}}{M_s}P_2^\text{cubic} = \frac{45\times 3^{1/4} \pi^{3/2}}{1024\, \Gamma
   \left(\frac{9}{4}\right)^2} \frac{(\op\bar r)^7}{ (\op\rv)^{3/2}} \, .
\end{equation}

The analytic expression~\eqref{powerradiated} is computed using the outgoing power for the perturbations of the field about a spherically symmetric background which accounts for the Vainshtein screening due to the monopole. This power is obtained from integrating $t_{0r}^\pi$, where $t_{\mu\nu}^{\pi}$ is the stress-energy tensor for the Galileon perturbations (see, e.g. \cite{Dar:2018dra})
\begin{align}
	t_{0r}^{\pi}=\frac32\left(1+\frac{4}{3\Lambda^{3}}\frac{E}{r}\right)\partial_{t}\pi\partial_{r}\pi.
	\label{approxpowernum}
\end{align}
Provided the power is computed by integrating over a sphere much larger than the Vainshtein radius, this should provide a reliable estimate of the nonlinear power. 

In practice, we calculate the power by defining a sphere of radius $r_* = \epsilon \bar{r}$, where $\epsilon = 22.5$, which is somewhat larger than $\rv = 20\bar{r}$ but less than half the size of the box, $L/2 = 50\bar{r}$.  Unlike the analysis of \cite{Dar:2018dra}, we choose to evaluate the radial flux on a set of points defined by the {\sc HEALPIX}\footnote{http://healpix.sourceforge.net} standard.  The values of $\pi$, $\dot{\pi}$ and $\partial_r \pi$ are calculated at all points over this sphere, {\sl even though they are not grid-points}, by doing a tri-linear interpolation.  Using this process allows us to (i) have assurances that the points are approximately equally-weighted when integrating over the sphere, and (ii) use efficient methods, provided by {\sc Healpy}, to decompose the fields onto spherical harmonics.

We use different software for each of the sections below; however all are based on GABE \cite{Child:2013ria}--a verified numerical tool for studying scalar fields.  While the numerical methods (and hardware) will vary from case to case, we will use the same fiducial physical system and numerical parameters, such as box-size, $L = 2.5 \rv = 50\bar{r}$ and number of points along each side, $N=384$.  In each of the simulations there will be a buffer of points--we generally take this buffer to the 6 grid-points nearest to any boundary-- around the boundary in which the field will not evolve according to the \eqn\eqref{eqn:eom}, but rather will evolve according to outward-going wave boundary conditions.  For the $\pi$ field, using an assumption that the non-linear terms are negligible at the boundary, this means
\begin{equation}
\label{eqn:boundary}
\dot{\pi} = -\frac{\pi}{r} - \partial_r \pi.
\end{equation}
While this assumption works very well for massless, Klein-Gordon scalar fields, it remains one of the greatest challenges to successfully simulating our systems.

In each simulation, we take the source to be two, rotating gaussian sources,
    \begin{equation}
        J = A \left(e^{-\left({\vec{r}_+}^{\,\prg}(t)/\sigma_\prg\right)^2}+e^{-\left({\vec{r}_-}^{\,\prg}(t)/\sigma_\prg\right)^2}\right),
    \end{equation}
where $ {\vec{r}_\pm}^{\,\prg} (t) = \left(x_\prg\pm \cos\left(\Omega_\prg t_\prg\right), y_\prg \pm \sin\left(\Omega_\prg t_\prg\right), z_\prg\right)$ and the constant 
\begin{equation}
A = \frac{2\sqrt{2}}{3\pi} \frac{\Omega \bar{r}}{\sigma_\prg^3} 
\end{equation}
is set such so that the total mass of the system is $M_{\rm s} = \int d^3x\, \rho = \int d^3 \,T$.  For our fiducial model, we will take $L = 50\bar{r}$, $N^= 384^3$ and will use a `standard' timestep $dt_{\rm fid} = \beta^{-1}\bar{r}/6400 \approx 0.003125 \bar{r}.$

\subsection{An active low-pass filter}
\label{ALPF} 

To recover and go beyond the analysis of \cite{Dar:2018dra}, we begin by attempting to simulate the fully-nonlinear system \eqref{eqn:eom}, using just one degree of freedom.

In this system, terms that involve products of second-derivatives are a particular numerical challenge.  In these cases, the accuracy of finite-differencing stencils (finite-approximations to calculating derivatives) is one of the main roadblocks for accurately evolving the field.  Accuracy can be gained by increasing the number of nearby-neighbors used to calculate these derivatives; however, there is a substantial run-time cost to that strategy.

Therefore, rather than using finite-derivative stencils, we use a spectral method, in which we take a Fourier transform of the scalar field, $\pi$, and its time-derivative, $\dot{\pi}$, at each step, and we then calculate the first- and second-derivatives of $\pi$ and $\dot\pi$ in momentum space before performing a set of inverse-Fourier transforms to recover the configuration-space derivatives.  This process gives excellent approximations to the derivatives of the field {\sl away from the boundary}.  Luckily, we only need to evolve the boundaries using \eqn\eqref{eqn:boundary} and we employ 2nd-order, inward finite-differencing stencils to calculate $\partial_r \pi$ in that region.

This process can be computationally expensive on a CPU, so we employ the GPU-accelerated version of GABE. This version, written in CUDA, maintains all the same structures of the original software, but is written such that the field evolution occurs on a GPU.  This acceleration is particularly useful for taking Fourier transforms and is ideally suited for our task.  In addition the GPU-accelerated version of GABE also uses a 4th order Runge-Kutta integrator, which, in principle allows us to use slightly larger time-steps as compared to GABE.  

However, the greatest benefit to using the GPU accelerated version of GABE is the ability to quickly apply a low-pass filter on the field (or on its derivatives) and thereby to actively remove any high-frequency modes.  In practice, we found that applying a low-pass filter,
\begin{equation}
F(k) = \frac{1}{2}\tanh^{-1}\left(\frac{1}{10}\left[\frac{N}{2}-\frac{k^2}{dk^2}\right]\right)+\frac{1}{2},
\end{equation}
to $\dot{\pi}$ at the end of every Runge-Kutta step, as well as to the Fourier-transform of mixed-spatial derivatives, gave excellent stability without the need to apply additional filters.  This filter is designed to cut-off power in modes larger than the one-dimensional Nyquist Frequency to $k_{\rm 1DN} = dk\,N/2 = \pi N/L$, where $dk = 2\pi/L$ is the standard unit for discrete Fourier Transforms, .

In order to achieve stability, however, one needs to employ slow `turn-on' strategies like those used in \cite{Dar:2018dra}.  In this scheme, 
\begin{equation}
\label{eqn:eom-pr}
\Box_\prg\pi_\prg + f_3(t_\prg)\kappa \left(  (\Box_\prg\pi_\prg)^{2} - (\partial^\prg_\mu\partial^\prg_\nu\pi_\prg)^2 \right) = f_1(t_\prg)J_\prg
\end{equation}
where $f_1$ and $f_3$ are window functions that start at zero `ramp-up' to unity,
\begin{align}
f_3(t_\prg) &= \frac{\tanh(.015(t_\prg - 350.)) + \tanh(5.25)}{1. + \tanh(5.25) - 0.01}\\
f_1(t_\prg) &= \frac{1}{2}\tanh^{-1}\left(\frac{1}{10}\left[t_\prg-25\right]\right)+\frac{1}{2}\,,
\end{align}
where the choice of smoothing parameter, $0.015$ is chosen as a reasonable numerical parameter and $5.25 = 350\times0.015$.

\begin{figure*}[tbp]
\centering
\includegraphics[width=0.45\textwidth]{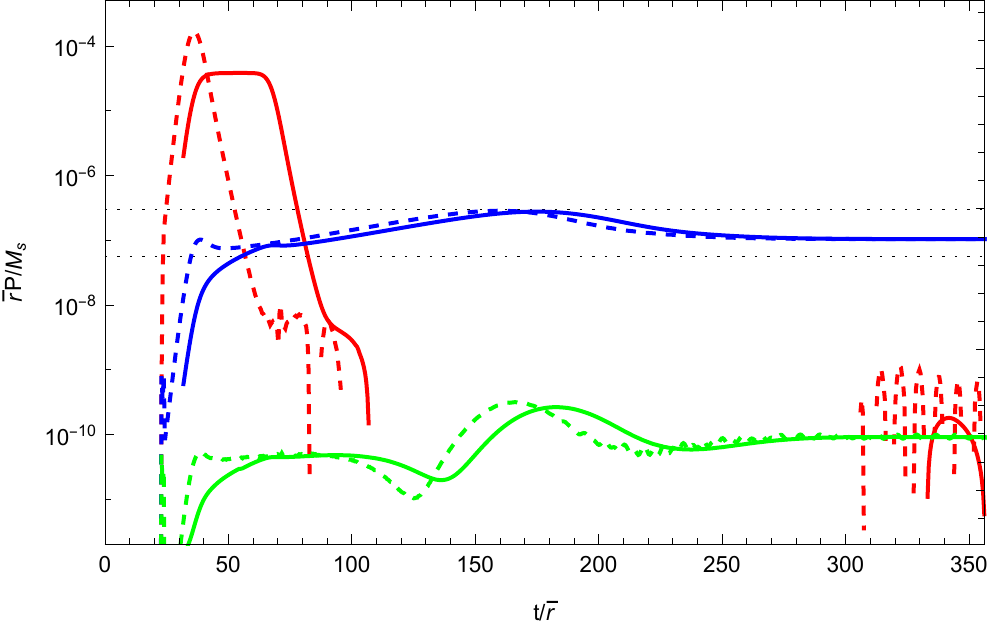}
\includegraphics[width=0.45\textwidth]{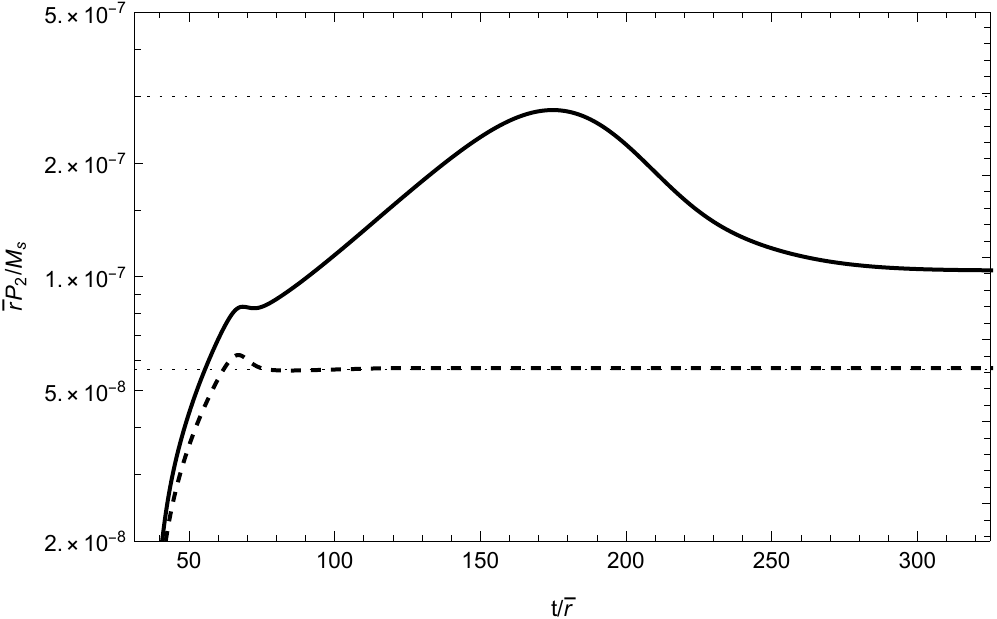}
\caption{(Left) The instantaneous (dotted) and period-averaged (solid) power emitted by the fiducial system for the monopole (red), quadrupole (blue) and $\ell=4$ mode (green). (Right) The quadrupole power emitted by the fiducial system (black line) employing an active low-pass filter (as described in Section \ref{ALPF}) as well as the quadrupole power emitted by the Klein-Gordon system. On both plots the lower dotted black line shows the analytic expectation for a Klein-Gordon Field and the higher dotted black line shows the analytic expectation for the fully non-linear system, \eqn\eqref{eq:powercomp}. Note that the vertical axis scale varies between the two plots.\label{ALPFfig}}
 \end{figure*}

Figure \ref{ALPFfig} shows the quadrupole power emitted in this system as a function of time.  The ratio of the late-time quadrupole power in the full system compared to a Klein-Gordon system is 1.81, which is consistent with the values seen in\cite{Dar:2018dra}.

In Figure \ref{ALPFfig} (and in following figures) for simplicity we plot the quantity  
\begin{align}
	\bar{P} \equiv \frac32 \int d\Omega \, r^2 \partial_{t}\pi\partial_{r}\pi \ ,
	\label{numericalpowercompare}
\end{align}
which is related to the analytic expression~\eqref{powerradiated} for the power radiated by 
\begin{equation}
	\bar{P}=P_2^\text{cubic} \left(1+\frac{4}{3\Lambda^{3}}\frac{E}{r}\right)^{-1} \approx 3\times 10^{-7}\frac{M_s}{\bar{r}} \, \\
	\label{eq:powercomp}
\end{equation}
which is useful when comparing multiple methods.

\subsection{Full Auxiliary Field Method}
\label{aux15}

The second approach is to define {\sl auxiliary fields}, $A_\mu \equiv \partial_\mu \pi$ and $H_{\mu\nu} \equiv \left(\partial_\mu A_\mu + \partial_\nu A_\mu\right)/2$, for which the classical equations of motion describing their interactions are \eqref{Piequation},\eqref{Aequation},\eqref{Hequation}. When converting these equations to program units, only a single window function, $f_1(t_\prg)$ is now needed,
\begin{align}
&\Box_\prg \pi + \kappa \left(H_\prg^{\mu \nu}H^\prg_{\mu \nu} - \left({H_\prg}^\nu_\nu\right)^2\right) = f_1(t) J_\prg \label{Piequationnum}\\
&\Box_\prg A^\prg_\mu - \frac{1}{\tau} \partial^\prg_t A^\prg_\mu - M_\prg^2 A^\prg_\mu = - M_\prg^2 \partial^\prg_\mu \pi \label{Aequationnum}\\
&\Box_\prg H^\prg_{\mu\nu} - \frac{1}{\tau} \partial^\prg_t H^\prg_{\mu\nu} - M_\prg^2 H^\prg_{\mu\nu} = - \frac{M_\prg^2}{2}\left(\partial^\prg_\mu A^\prg_\nu + \partial^\prg_\nu A^\prg_\mu\right).\label{Hequationnum}
\end{align}
where $M_\prg = M \bar{r}/2$.

These massive auxiliary fields cannot use the same massless outgoing wave boundary conditions that we described earlier for the $\pi$ field. 
Instead, we enforce the constraint equations for $A_\mu$ and $H_{\mu \nu}$ given in \eqref{Aapprox} and \eqref{Happrox} when the waves reach the buffer (defined as $N/64$ where $N$ is the size of the box).
Using these relaxed constraints, we have been able to achieve numerical stability regardless of when the source `turned-on'. 

In addition to the fiducial tests, we also make a single comparison to a larger, $N^3=512^3$ simulation for $M \bar{r}= 10$.  This run will be important as a comparison where we keep the grid-spacing, $dx = L/N$, constant therefore moving the boundary away from the source without changing the range of high-frequency modes in our system.  This run is of particularly importance in diagnosing the limitations of Auxiliary fields as a numerical scheme.

\begin{figure}[ht]
\centering
\includegraphics[width=0.45\textwidth]{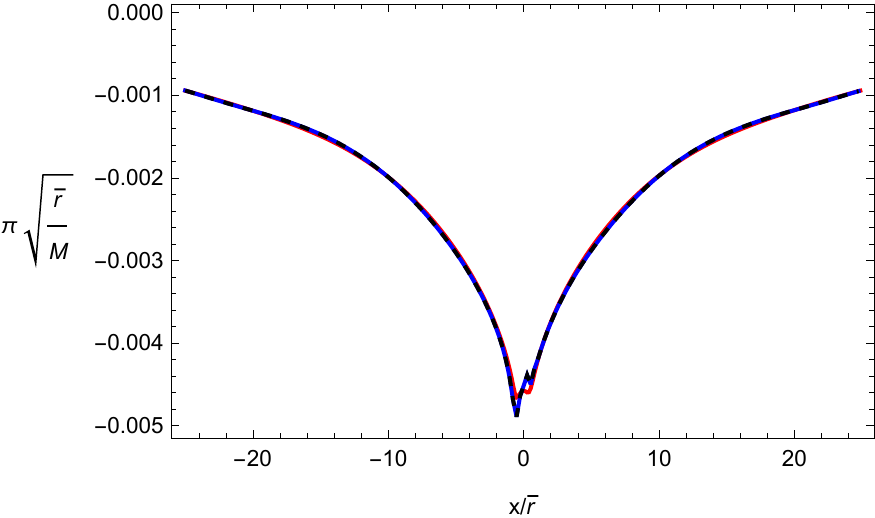}
\caption{The profile of the $\pi$ field along the $x$-axis for the active low-pass filter system (black, dashed), as well as the fiducial $N^3=384^3$ (blue) and larger, $N^3=512^3$ (red), simulations using Auxiliary fields.  \label{profile}}
\end{figure}

We can compare the results of this system to the full system in a couple of different ways.  As a first comparison, we look at the profile of the $\pi$ field along a line in the equatorial plane of the binary system (take here to be the $x$-axis).  Figure \ref{profile} shows excellent agreement between the active low-pass filter and the Auxiliary field methods, particularly at $N^3=384^3$, as well as agreement with the larger, $N^3=512^3$ simulation.

\begin{figure*}[ht]
\centering
\includegraphics[width=0.45\textwidth]{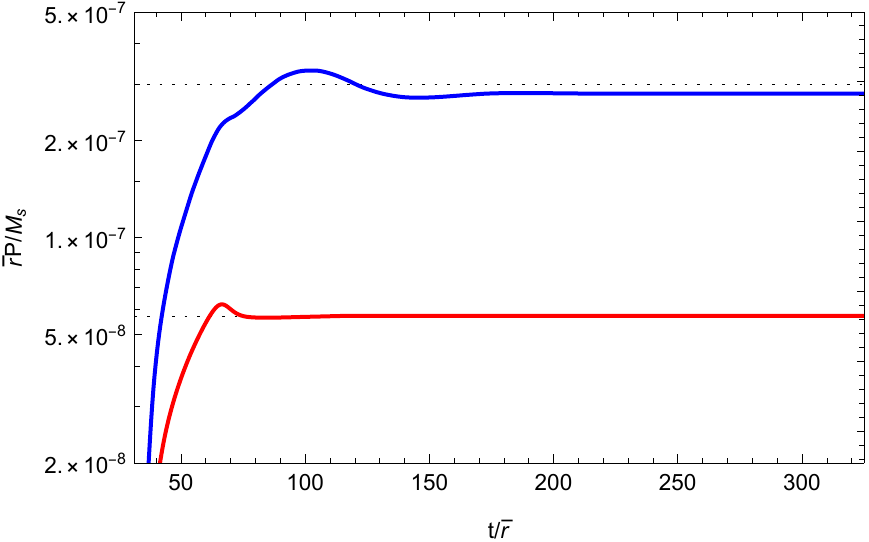}
\includegraphics[width=0.45\textwidth]{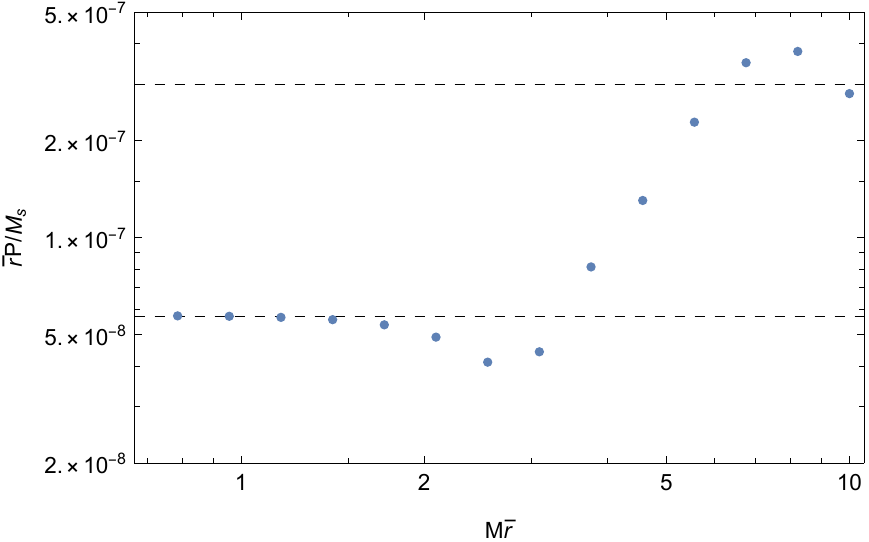}
\caption{(Left) The quadrupole power emitted by the fiducial system using auxiliary fields (as described in Section \ref{aux15}) for $M\bar{r}\approx 0.8$ (red) and $M\bar{r} \approx 10$ (blue).  (Right) The late-time quadrupole power emitted by the fiducial system using auxiliary fields for different values of $M$.  In both panels, the lower dotted black line shows the analytic expectation for a Klein-Gordon Field and the higher dotted black line shows the analytic expectation for the fully non-linear system.  \label{powercomp}}
\end{figure*}

Next, we can look to the multipole power radiated in the system by the $\pi$-field.  Figure~\ref{powercomp} shows the period-averaged power in the quadrupole for a fairly low, $M\bar{r} \approx 0.8$, approximately Klein-Gordon simulation as well as the largest mass, $M\bar{r} = 10$, that reached equilibrium.  Figure~\ref{powercomp} also shows the parametric dependence of the final, quadrupolar power versus $M\bar{r}$.  The progression from near-Klein-Gordon to approaching the full, nonlinear system occurs as $M$ transitions from a small to large number compared to one.  Below we comment on the limitations of our numerical system to go to higher values of $M$, however, we anticipate that the trend shown on the right panel of Figure~\ref{powercomp} would continue until it matches Figure~\ref{ALPFfig}.

\begin{figure*}[htbp]
\centering
\includegraphics[width=0.45\textwidth]{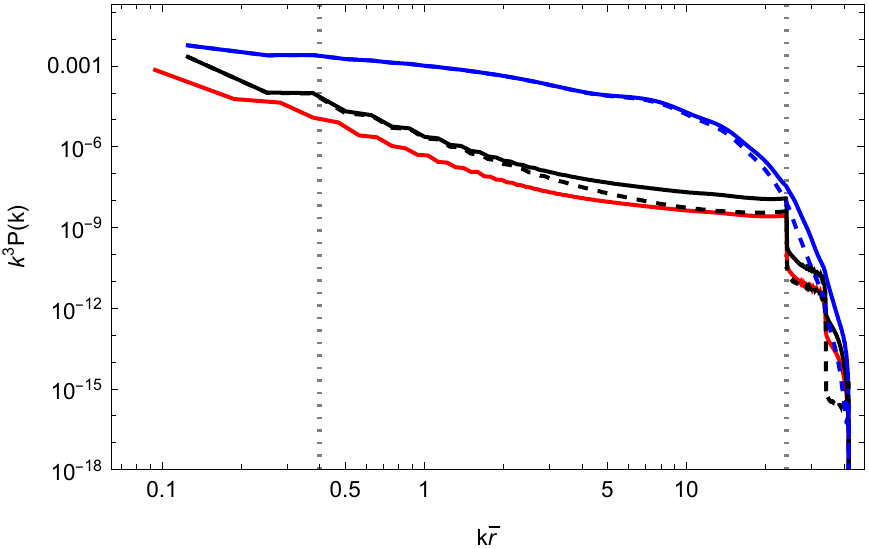}
\includegraphics[width=0.45\textwidth]{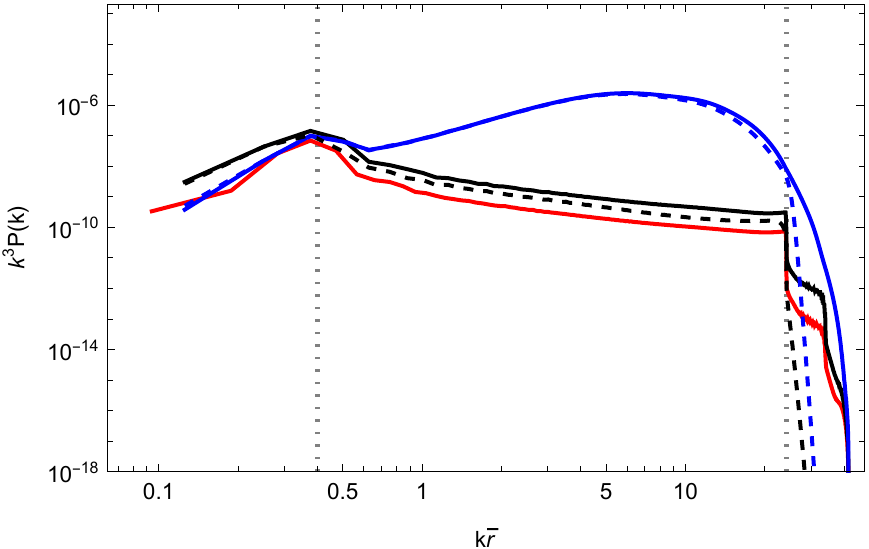}
\caption{The dimensionless power spectrum of the $\pi$ field after the system has reached equilibrium (Left) and the dimensionless power spectrum of $\dot{\pi}$ field after the system has reached equilibrium (right).  The blue lines show the results of a near-Klein-Gordon field using Auxiliary fields ($M\bar{r} \approx 0.8$, solid lines) and a true Klein-Gordon field using the active low-pass filter (dashed lines).  The single solid red line is for an $N^3=512^3$ simulation of Auxiliary fields with a larger, $L=66.66\bar{r}$, box.  In both plots the leftmost vertical dashed line corresponds to the frequency, $2\Omega\bar{r}$, where one would expect to see quadrupole power from a binary system and the right vertical dashed line corresponds to the one-dimensional Nyquist frequency, $\pi N/L$ \label{power}}
\end{figure*}

In addition to considering the period-averaged power, we can also look for consistency in the power spectra of the $\pi$ field and its derivative.  Figure~\ref{power} compares the dimensionless power spectra of the $\pi$ field, as well as its time-derivative, $\dot{\pi}$.  These plots show exceptional consistency between our different methods as a mode-by-mode comparison.  This figure also shows the effect of the nonlinear terms on the system; the Klein-Gordon (or near Klein-Gordon) simulations have significantly more power on smaller scales which is suppressed as the nonlinear terms become important.

\begin{figure}[htbp]
\centering
\includegraphics[width=0.45\textwidth]{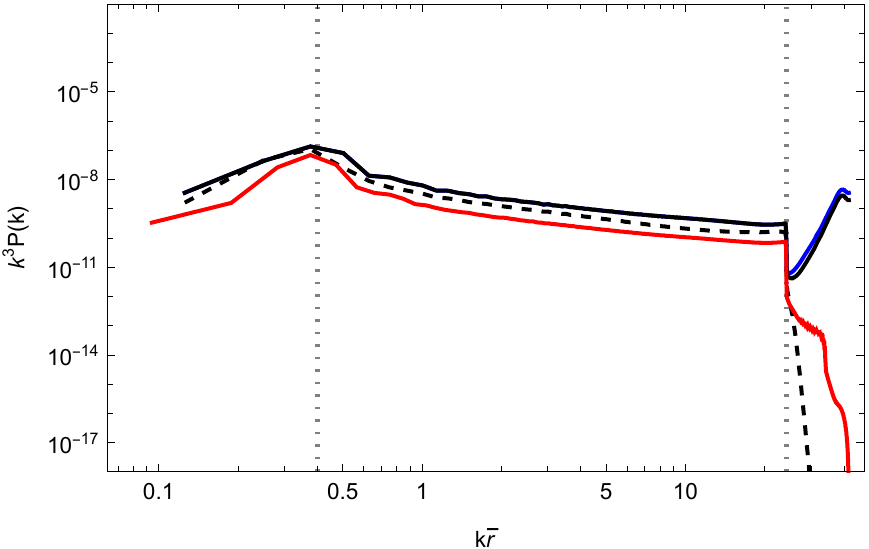}
\caption{The dimensionless power spectrum of $\dot{\pi}$ field near the final time.  The solid curves represent two $N^3=384^3$ simulations using Auxiliary fields and $M\bar{r} = 10$ with different timesteps, $dt_{\rm fid}$ (blue) and $dt_{\rm fid}/2$ (black), the dashed black curve is a simulation using an active low-pass filter and the red curve is a $N^3=512^3$ simulation using Auxiliary fields ($M\bar{r} = 10$) with a larger, $L=66.667\bar{r}$, box, a fiducial $dx$ and $dt = dt_{\rm fid}/2$. The leftmost vertical dashed line corresponds to the frequency, $\Omega\bar{r}$, and the right vertical dashed line corresponds to the one-dimensional Nyquist frequency, $\pi N/L$.  The slices are all taken at the same late-time when the $N^3=384^3$ simulations are about to crash. \label{powerbreak}}
\end{figure}

One of the issues we encountered while simulating this model was that the code would crash as we increased the mass of the Auxiliary fields, $M$.  For our simulations, long-term stability became intractable around $M\bar{r} \approx 10$.  For the specific borderline case of $M\bar{r} = 10$, our fiducial model was able to achieve stability for many orbits of the system; however, after some time high-frequency modes are excited and the code becomes unstable.  This instability does not seem to arise from a problem with the dynamics of the system, rather, it emerges as a consequence of our boundary conditions.  In the boundary, we calculate the derivatives of the Auxiliary fields assuming that the constraints are satisfied and \eqn\eqref{eqn:boundary}.  This is a good approximation if (1) we are sufficiently far away from the source such that the $\pi$ field is Klein-Gordon and (2) the constraints are satisfied exactly.  For values of $M\bar{r} > 5$ we seem to violate these assumptions.  To demonstrate, we look at our marginal, $M\bar{r}= 10$, case, and test whether the instability is a consequence of numerical instability (by reducing the time-step) or a result of the boundary conditions (by keeping $dx$ the same, but increasing resolution to send the boundary further away from the source).  Figure \ref{powerbreak} shows that the simulations are not stabilized by increasing time resolution (which would indicate that we're not numerically resolving the problem well); however, the system remains stable for much longer if the boundary is moved away from the source.

\subsection{Restricted Auxiliary Field Method}
\label{aux10}

In addition to the above described UV completion, we also numerically explore a partial UV completion which is obtained from the system \eqref{Piequation},\eqref{Aequation},\eqref{Hequation} by taking the scaling limit $M \rightarrow \infty$ for fixed 
\begin{equation}
\hat \tau= \frac{1}{\tau M}  \, .
\label{defoftauhat}
\end{equation}
In this limit, the equation of motion for the $\pi$ field remains the same, however those for the additional fields can be reduced to second order equations of motion for the ten auxiliary fields, $H_{\mu\nu}$, given by
\begin{equation}
\left( 1+  \hat \tau \partial_t \right)^2 H_{\mu\nu}= \partial_{\mu}\partial_{\nu} \pi \, .
\label{restraux}
\end{equation}
or more explicitly
\begin{equation}
\ddot{H}_{\mu \nu} = \frac{1}{\hat{\tau}^2} \left(\partial_\mu \partial_\nu \pi\right) -  \frac{2}{\hat{\tau}} \dot{H}_{\mu \nu} - \frac{1}{\hat{\tau}^2} H_{\mu \nu} \, .
\end{equation}
This restricted system is similar in spirit to the approach taken in \cite{Cayuso:2017iqc,Allwright:2018rut} based on the M\"uller-Israel-Stewart formulation \cite{muller1967paradoxon,ISRAEL1976213,ISRAEL1979341,ISRAEL1976310} which has recently been successfully applied to effective field theories of gravity in \cite{Cayuso:2020lca} (for related work on cubic Horndeski theories see \cite{Figueras:2021abd}).
In this approximation, as with the fully UV complete system, we employed outward-going boundary conditions on the $\pi$ field.
In contrast to the full system of Auxilliary fields, however, we did not need to enforce any boundary conditions on the $H_{\mu \nu}$ fields since these restricted Auxiliary fields are not propagating degrees of freedom and neither the equations of motion for $\pi$ nor for $H_{ \mu \nu}$ depend on derivatives of $H_{ \mu \nu}$.  Given this, we only need to define $H_{\mu \nu}$ in the bulk.

\begin{figure}[htbp]
\centering
\includegraphics[width=0.45\textwidth]{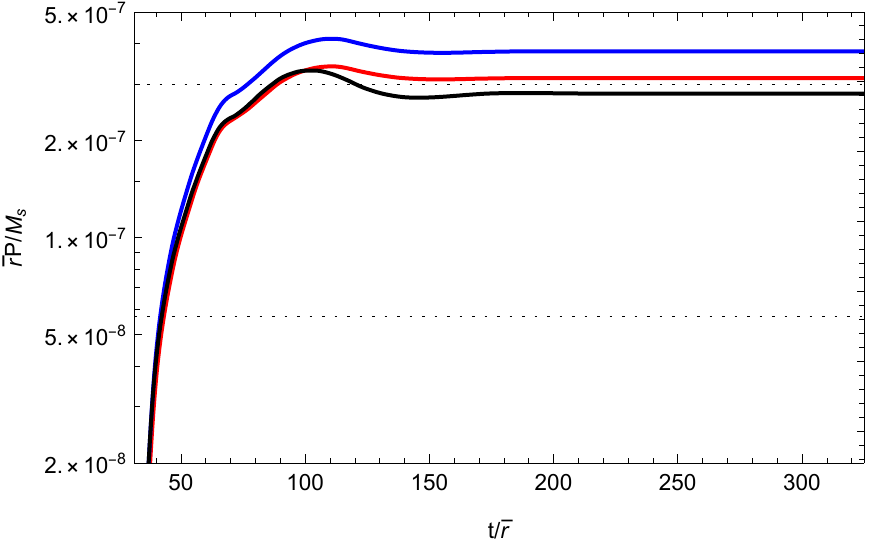}
\caption{The period-averaged quadrupole power using the system of restricted auxiliary fields, \eqn\eqref{restraux}, and $\hat{\tau} \approx 5.9$ (red) (in program units) with the full auxiliary fields for $M\bar{r} \approx 8.2$ (blue) and $M \bar{r} \approx 10$ (black).  \label{110vs17}}
\end{figure}

To compare it to the first system, we simulate this system using numerical parameters comparable to the largest stable value of $M\bar{r} = 8.22$---calculating $\hat{\tau}$ from \eqn\eqref{defoftauhat}.  Figure~\ref{110vs17} shows a comparison of the period-averaged quadruple power.  

\section{Discussion}
Effective field theories inevitably involve derivative interactions, the effects of which can have important and interesting implications in a number of settings, particularly in gravitational physics and cosmology. While it is well-understood how to analytically deal with the subtleties of solving the resulting equations of motions, significant problems can arise in numerical implementations. This fact has seriously hampered progress in understanding the detailed predictions of large classes of theories that have received much recent attention.

In this paper we have developed, compared, and contrasted three ways of dealing with this problem in numerical implementations of such theories. The first approach is to employ a low-pass filter to tame the UV modes. The second approach is to construct an example of a ``UV-completion" of the equations of motion, involving auxiliary fields that constitute new propagating degrees of freedom. The effect of these fields is to render the full system of equations formally well-posed (the system is hyperbolic for all degrees of freedom, and the characteristic speed is unity for all modes), but also to ensure that the IR behavior lies in the same universality class as the original set of equations. The third approach is a restricted ``UV-completion", also using  auxiliary fields but without introducing new propagating degrees of freedom. The key point here is that we posit equations of motion that, while remaining second order, now involve damping terms to again tame the UV behavior. 

Explicitly, we have simulated an orbiting two body system, and determined the power spectrum of scalar radiation of relevance for example to binary pulsars in common examples of modified theories of gravity. We have demonstrated that for the same initial data, all three methods reproduce the same long wavelength physics with the expected errors of the numerical simulations. In the case of the low pass filter, both the source and interactions need to be turned on slowly in order to maintain numerical stability. On the other hand, both of the UV completions are found to be under better control, allowing the interactions to be turned on at the initial timestep. These results parallel those of \cite{Babichev:2017lrx} and more recently \cite{Bezares:2021yek,Lara:2021piy}, which consider UV completions of theories with kinetic screening along the lines of \cite{Tolley:2009fg,Elder:2014fea,Solomon:2020viz}.

One remaining technical issue that prevents us from treating large hierarchies of scale (large $M$) is that our treatment of the boundary conditions for the massive degrees of freedom is in tension with the damping of the bulk degrees of freedom. This problem arises because of a known issue with imposing boundary conditions in real space for massive fields (see, for example,~\cite{Honda:2000gv}). A better treatment of boundary conditions should remove this issue.

Our hope is that the techniques described in this paper will be of direct use to those wishing to simulate generic effective field theories, including known difficult examples such as Galileons, massive gravity, and the effects of higher-curvature corrections in gravity.


\begin{acknowledgments}

AJT and MT would like to thank Luis Lehner for extremely useful discussions and for a careful reading of the manuscript.  We thank the National Science Foundation, and the Kenyon College Department of Physics for providing the hardware used to carry out these simulations. M.G. and J. T. G. are supported by the National Science Foundation, Grant No. PHY-2013718. AJT is supported by STFC grant ST/T000791/1.  AJT thanks the Royal Society for support at ICL through a Wolfson Research Merit Award. MT is supported in part by US Department of Energy (HEP) Award DE-SC0013528.  Some of the results in this paper have been derived using the healpy and HEALPix packages \cite{Zonca2019,2005ApJ...622..759G}.

\end{acknowledgments}

\bibliographystyle{apsrev4-1} 
\bibliography{references.bib}

\end{document}